\documentclass[preprint]{aastex}
\shorttitle{Intercahnge reconnection assoicated with a confined filament eruption}
\slugcomment{Submitted to ApJ}
\shortauthors{Zheng et al.}
\begin{document}

\title{Interchange reconnection associated with a confined filament eruption: Implications for the source of transient cold-dense plasma in solar winds}
\author{Ruisheng Zheng$^{1}$, Yao Chen$^{1}$, Bing Wang$^{1}$, and Gang Li$^{2}$}
\affil{$^{1}$Shandong Provincial Key Laboratory of Optical Astronomy and Solar-Terrestrial Environment, and Institute of Space Sciences, Shandong University, 264209 Weihai, China; ruishengzheng@sdu.edu.cn\\
 $^{2}$Department of Physics and CSPAR, University of Alabama in Huntsville, Huntsville, AL 35899, USA\\}

\begin{abstract}
The cold-dense plasma is occasionally detected in the solar wind with in situ data, but the source of the cold-dense plasma remains illusive. Interchange reconnections (IRs) between closed fields and nearby open fields are well known to contribute to the formation of solar winds. We present a confined filament eruption associated with a puff-like coronal mass ejection (CME) on 2014 December 24. The filament underwent successive activations and finally erupted, due to continuous magnetic flux cancellations and emergences. The confined erupting filament showed a clear untwist motion, and most of the filament material fell back. During the eruption, some tiny blobs escaped from the confined filament body, along newly-formed open field lines rooted around the south end of the filament, and some bright plasma flowed from the north end of the filament to remote sites at nearby open fields. The newly-formed open field lines shifted southward with multiple branches. The puff-like CME also showed multiple bright fronts and a clear southward shift. All the results indicate an intermittent IR existed between closed fields of the confined erupting filament and nearby open fields, which released a portion of filament material (blobs) to form the puff-like CME. We suggest that the IR provides a possible source of cold-dense plasma in the solar wind.
\end{abstract}


\section{Introduction}
The generation of solar winds is one of the active research topics, and the existence of solar winds owes to the extended hot corona (Edmonson et al. 2012). Though it is rare, the transient cold-dense plasma is sometimes detected in the solar wind with in situ observations. Most of the cases are associated with interplanetary coronal mass ejections (ICMEs) or magnetic clouds (MCs), and the cold-dense plasma is identified as the ICME core consisting of filament material that shows a low proton temperature, a high proton density, a flux rope, and the ions with low charge states (Yao et al. 2010; Lepri \& Zurbuchen 2010; Song et al. 2017). On the other hand, for normal solar winds not associated with ICMEs/MCs, the origin of their cold-dense plasma are not resolved.

It has been suggested that the release of the filament material along open field lines may account for the origin of the transient cold-dense plasma of solar winds. Filaments, also known as prominences when observed in the solar limb, consist of cold-dense plasma embedded in the hot corona (Mackay et al. 2010). Filaments are usually formed along the magnetic polarity inversion lines (PILs; Martin 1998; Berger et al. 2008; Ning et al. 2009). It is not clear how these filament material are released into the open field lines along which the solar wind propagates outward.

The interchange reconnection (IR; Crooker et al. 2002), between open and closed fields, is often thought to occur at the apex of streamers or pseudo-streamers near a null point (Wang et al. 2012). In a model of turbulent corona, Rappazzo et al. (2012) suggested that the IR may happen everywhere along the interface region between the nearby coronal hole and closed loops threaded by a strong unipolar magnetic field. The IR allows closed field lines to open, which can release the coronal loop plasma into the interplanetary. IRs between erupting filaments and coronal holes or pseudo-streamers have been reported (Zhu et al. 2014; Yang et al. 2015). Hence, the IR is a primary mechanism to contributes mass, heat, and momentum to solar winds, especially the highly variable (in densities, temperatures, velocities, etc) slow solar wind (Wang et al. 1998; Fisk et al. 1999; Crooker et al. 2002; Antiochos et al. 2007; Edmondson et al. 2009; Masson et al. 2012).

In this paper, we report a likely IR between closed field lines of a confined erupting filament and nearby open field lines. The IR released a portion of filament material along newly-formed open filed lines, which shed light on the possible source of the cold-dense plasma of solar winds. The paper is organized as follows. Section 2 describes the observations used in this work; The main results are showed in Section 3; We give conclusions and discussion in Section 4.

\section{Observations and Data Analysis}
The filament eruption occurred at the NOAA active region (AR) 12421 on 2014 December 24, which was associated with a C3.7 flare. To study the filament evolution, we employ the observations from the Atmospheric Imaging Assembly (AIA: Lemen et al. 2012) onboard the {\it Solar Dynamics Observatory (SDO}: Pesnell et al. 2012) combining the H$\alpha$ filtergrams from the New Vacuum Solar Telescope (NVST; Liu et al. 2014) in the Fuxian Solar Observatory of China and from the Global Oscillation Network Group (GONG) of the National Solar Observatory. The AIA observes the full disk (4096~$\times$ 4096 pixels) of the Sun and up to 0.5 $R_{sun}$ above the limb in ten EUV and UV wavelengths, with a pixel resolution of 0$\farcs$6 and a cadence of 12 s. The NVST data used in this study were obtained in H$\alpha$ 6562.8~{\AA} from 08:11:22 to 08:51:56 UT on 2014 December 23, with a pixel size of 0$\farcs$163 and a field of view (FOV) of 151$\farcs$ $\times$ 151$\farcs$, covering AR 12421. The H$\alpha$ images of GONG are at 6563~{\AA} with a spatial resolution of 1$\arcsec$ and a cadence of $\sim$1 minute (Harvey et al. 2011).

Magnetograms from the Helioseismic and Magnetic Imager (HMI: Scherrer et al. 2012), with a cadence of 45 s and pixel scale of 0$\farcs$6, are used to check the magnetic field evolution of the source region. The CME is detected by the Large Angle and Spectrometric Coronagraph (LASCO; Brueckner et al. 1995) C2. All the SDO data are downloaded as full disk images, which allows to remove most of the de-rotation artefacts. Then, the images are de-rotated with a reference image at 02:00 UT on December 24, using the SolarSoft/IDL routine DROT\_MAP. The kinematics of the filament branches and associated mass flow are analyzed with the time-slice approach. The speeds are determined by linear fits, with error bars given by the measurement uncertainty of 4 pixel ($\sim$1.74 Mm) for AIA data. We also utilize the Potential Field Source Surface (PFSS; Schrijver {\&} De Rosa 2003) model to extrapolate the large-scale magnetic field topology.

\section{Results}
\subsection{Confined Filament Eruption}
Figure 1 shows the magnetic filed evolution of AR 12421 in HMI magnetograms and intensity maps. The AR (the white box in panel a) is close to the west limb, and there is a large patch of negative polarity to the southwest of the AR (the white arrow in panel a). In the intensity maps (panels b-c), the AR clearly consists of the west leading main sunspots and the east following small spots, and the following spots become fainter and fainter (white boxes), indicating the magnetic flux cancellation. Along the eastern part of the PIL, there are some magnetic cancellation sites (red arrows in panels d-f) and some emergence locations (blue arrows in panel g-h), which is very distinct in the attached animation of magnetograms. In the plot of the total flux change (panel i) for the magnetic field in the east of the AR (white boxes in panels b-h). For both the positive and negative fluxes, there has a decreasing trend with some bumps. Note that, for the positive flux, a small bump (the arrow) appeared just before the eruption onset (the vertical line). It indicates that the eruption is closely associated with the abrupt emergence of the positive flux.

As a result of the continuous magnetic activities, the filament underwent some activations before the eruption, shown in H$\alpha$ filtergrams of NVST and GONG, and in AIA 304~{\AA}(Figure 2). The activation at $\sim$08:10 UT on December 23, clearly captured by the NVST H$\alpha$ filtergrams (see the attached animation). The fine structures of the filament are shown obvious, and the activated plasma moved along the twisted strands (arrows in panel a). Superposed with the HMI magnetogram, the filament clearly lay along the eastern part of the PIL, and its south and north footpoints rooted in negative and positive polarities, respectively (arrows in panels b-c). About 20 minutes later, the activated filament returned to the peaceful condition (panel c). Other two activations are simply displayed in H$\alpha$ filtergrams of GONG, and the brightenings were diffuse (black arrows in panels d-e). The filament survived during the successive activations (panel g). At $\sim$02:30 UT on December 24, there appeared again some brightenings near the filament center (arrows in panels f and h), and then the filament began to erupt (panel i).

Figure 3 displays clearly the filament eruption in GONG H$\alpha$ filtergrams (panels a-c), in AIA 304 (panels d-f and i) and 335~{\AA} (panels g-h) images. The filament experienced a clear untwist motion, but it only rose a short distance and stopped (see the attached animation). Therefore, the eruption is apparently confined (left and middle panels). After the eruption, most of the filament material was trapped in the corona (the white arrow in panel c), and a thin filament existed clearly along the PIL (the black arrow in panel c). Interestingly, some plasma moved to a remote site to the southwest of the AR (black arrows in panel f), and a section of a quasi-circular flare ribbon bounding the remote site to the north end of the filament (panel i). It likely implies that some magnetic reconnection occurred beneath a coronal null point, and the large quasi-circular separatrix footpoints of the null point produces the quasi-circular flare ribbons (Masson et al. 2012; Filippov 2015). Moreover, some blobs clearly escaped from the confined filament body (the white arrow in panel f).

\subsection{Interchange Reconnection}

Next, we focus on the mass release from the trapped filament, better shown in the images rotated anticlockwise in 90$^\circ$ (Figure 4). The blob escaped in a curved trajectory (better seen in the accompanying animation), and its speed was $\sim$110 km s$^{-1}$, estimated by the successive positions (black arrows) in the top panels. Intriguingly, it was clear for the opening of some filament threads where the blobs escaped (the white arrow in panel b), and there appeared some credible open field lines (white arrows in panels c-e). The blobs and newly-formed open field lines are likely the indicators of the IR, and the newly-formed open field lines (L1) and their footpoints were distinct in 193~{\AA} (white arrows in panel f). Some minutes later, more open field lines formed (white arrows in panels g-h), which possibly suggests that the IR was slow and intermittent than that during a successful eruption. The blobs appeared to be rotating as it propagated outwards, which suggests that it was on its own untwisting flux tube. It is consistent with the recent mechanism producing such rotating blobs during intermittent IRs (Wyper et al. 2016). Note that L1 clearly shifted southward, supported by the black-white strips (white arrows in panels i-j). The southward movement of L1 along the S1 (black line in panel e) is shown in time-slice plots (panels k-l). The distinct V-type structure (the single black arrow in panel k) indicates that L1 first moved northward and then shifted southward, and the different branches (twin black arrows in panel k) represent different strands of L1 in different heights (panel k). The northward and southward speeds of the L1 were $\sim$79.2 and $\sim$142.4 km s$^{-1}$, respectively. In addition, some new closed loops (L2) formed during the IR, and L2 and their footpoints were also obvious (black arrows in panels g-j).

The evolution of L2 is shown in the AIA running-difference images in Figure 5. The moving plasma (the white arrow in panel a) appeared nearly simultaneously with the blob, and it likely flowed along the newly-formed L2. As a result, brightenings appeared at both ends of L2 (white arrows in panels b). Superposed with the HMI magnetogram, it is clear that the footpoints of L1 (the black arrow in panel b) rooted in the negative polarities around the south end of the filament, and the L2 connected with the positive polarities around the north end of the filament and the negative polarities (the arrow in Figure 1a) to the southwest of the AR. Note that L2 expanded at different height (black arrows in panel c), which was consistent with the intermittent IR. In addition, there were post-eruption loops straddling the PIL (twin white arrows in panel c). In AIA 304, 171, and 211{\AA}, L2 appeared simultaneously in lower and higher altitudes (black arrows in panels d-f). In 94 and 131~{\AA}, the lower branch of L2 was weak, and the higher branch of L2 was very faint (panels h-i). The weak brightenings around south footpoints of L2 were obvious in all AIA channels (white single arrows in panels c-i).

\subsection{CME}
Following the confined eruption and mass release, we found a CME in LASCO C2{\footnote{\url{http://cdaw.gsfc.nasa.gov/CME\_list/UNIVERSAL/2014\_12/yht/20141224.032405.w042n.v0664.p273g.yht}}} (Figure 6). The CME emerged between two streamers (panel a), and it first appeared in the FOV of LASCO C2 at 03:24 UT as a puff-like structure (panel b), nearly 35 minutes later than the blob onset. The central position angle of the CME was 255$^\circ$, and its median speed was 664 km s$^{-1}$. The CME experienced a southward shift during its propagation (panel c), and some new bright structures emerged behind the first bright front (arrows in panel d), which is consistent with the movement of the L1. The CME was much faster than the blob, which is likely as a result of the strong thermal pressure imbalance in the L1. In addition, the clear rotation and untwisting of the L1 also indicated magnetic tension that possibly drove plasma outwards  in the whip-like manner (Shibata \& Uchida 1986). Hence, the released mass/blobs were easily accelerated.

By examining the PFSS results of the AR magnetic field lines (panel e), there are open field lines (pink lines) rooting at negative polarities to the southwest of the AR. The filament erupted from the AR (the yellow arrow), and ran into nearby open field lines. It is possible that the IR occurs between the closed fields of untwisting filament and nearby open fields, which generates a CME taking off between the two streamers (the red arrow).

\section{Conclusions and Discussion}
Using the high-quality observations, we report a confined filament eruption associated with a puff-like CME on 2014 December 24. The close temporal relation indicates that a sequence of filament activations were likely resulted from continuous magnetic cancellations and emergences in the decaying AR. The existence of the IR between the closed fields of the confined erupting filament and the nearby open fields is supported by newly-formed open field lines (L1), newly-formed closed loops (L2), and the blobs and bright plasma moving along L1 and L2. The slow vertical movement of L1 and multiple branches of L1 and L2 likely indicate that the IR was intermittent during the confined filament eruption. The close temporal-spatial correlations and the similar southward movement suggest that the blobs likely finally evolved into the puff-like CME.

The scenario of the IR is shown in the schematic representation in Figure 7. The untwist and rising motion of the filament (the yellow thick rope) brings their closed field lines (deep blue) to meet with the oppositely aligned open field lines (pink), resulting in the IR. As a result of the IR, the newly-formed open field line (L1; red) rooting at negative polarities around the south end of the filament, and the newly-formed closed loops (L2; light blue) connect positive polarities around the north end of the filament and negative polarities of the open field to the southwest of the AR. On the other hand, the survived thin filament (the yellow thin rope) was possibly reformed or remained. The densities of the reconnecting closed and open field lines may differ by one to two orders in magnitude. Therefore, after the reconnection, L1 and L2 are in strong thermal/magnetic pressure imbalance between the part belonging to the former filament field lines and the section belonging to the former open field lines (e.g., Del Zanna et al. 2011). As a consequence, the blobs and bright plasma begin to move (white arrows along L1 and L2). During the IR, only a portion of closed filament field lines reconnected, and the more intact closed field lines (deep blue lines in panel b) restrained the erupting filament.

A small portion of the filament material (blobs) is released along newly-formed open fields, and formed a puff-like CME. We checked the in situ observations from {\it Advanced Composition Explorer (ACE)}, {\it Wind}, and {\it Solar TErrestrial RElations Observatory (STEREO)}, but there was no clear signal of the counterpart of the blobs in the interplanetary. We suppose that these plasma is associated with the magnetic structure of open field lines, not closed ICME/MC-like features. Their interplanetary counterpart may mimic the magnetic configuration of the normal solar wind. Due to the filament plasma is about 10-100 times colder and denser than surroundings, they will certainly experience some thermal-dynamical evolution during their propagation to the interplanetary space. Yet, it is reasonable that they remain similarly cold and dense, especially when the mixing with the background plasma is not very effective. Thus, the study provides a possible mechanism for the origin of the transient cold-dense plasma that is sometimes detected in the solar wind by spacecrafts (Yao et al. 2010; Lepri \& Zurbuchen 2010; Song et al. 2017).

\acknowledgments
The authors thank the anonymous referee for constructive comments. We gratefully acknowledge the usage of data from the SDO and SOHO spacecraft, and the ground-based GONG and NVST projects. This work is supported by grants NSFC 41331068, 11303101, and 11603013, Shandong Province Natural Science Foundation ZR2016AQ16, and Young Scholars Program of Shandong University, Weihai, 2016WHWLJH07.

\clearpage

\begin{figure}
\epsscale{0.9} \plotone{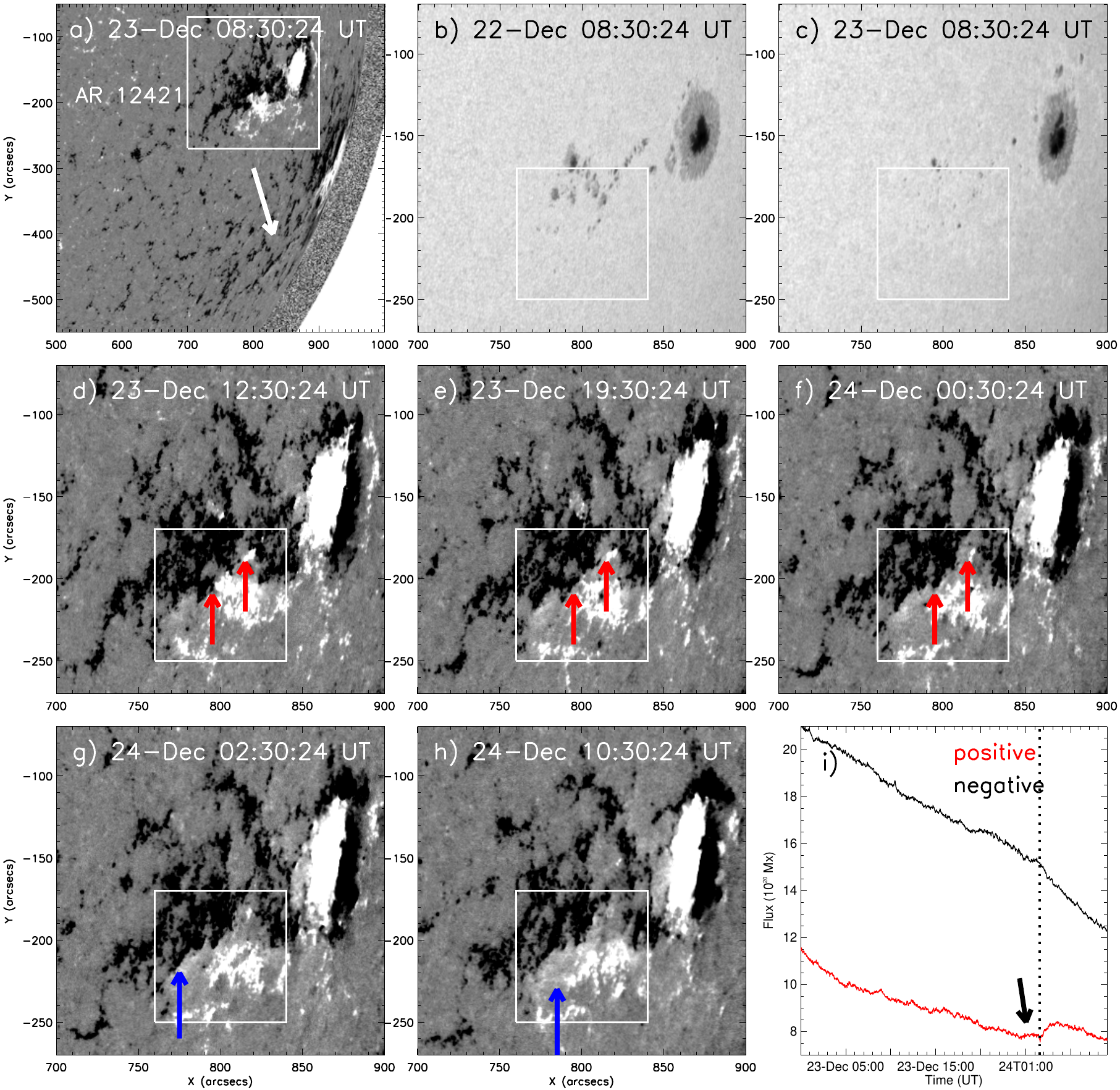}
\caption{(a) Overview of AR 12421 (the white box that represents the FOV of (b-h)) and a nearby open field (the white arrow) in a HMI magnetogram. (b-h) The evolution of AR 12421 in HMI intensity maps and magnetograms. The red and blue arrows indicate the sites of magnetic cancellation and emergence, respectively. (i) The changes of the total magnetic flux for the eastern part of the AR (the boxes in (b-h)). The black arrow indicates the positive flux emergence just before the eruption, and the dotted line marks the eruption onset.
\label{f1}}
\end{figure}

\clearpage

\begin{figure}
\epsscale{0.9}
\plotone{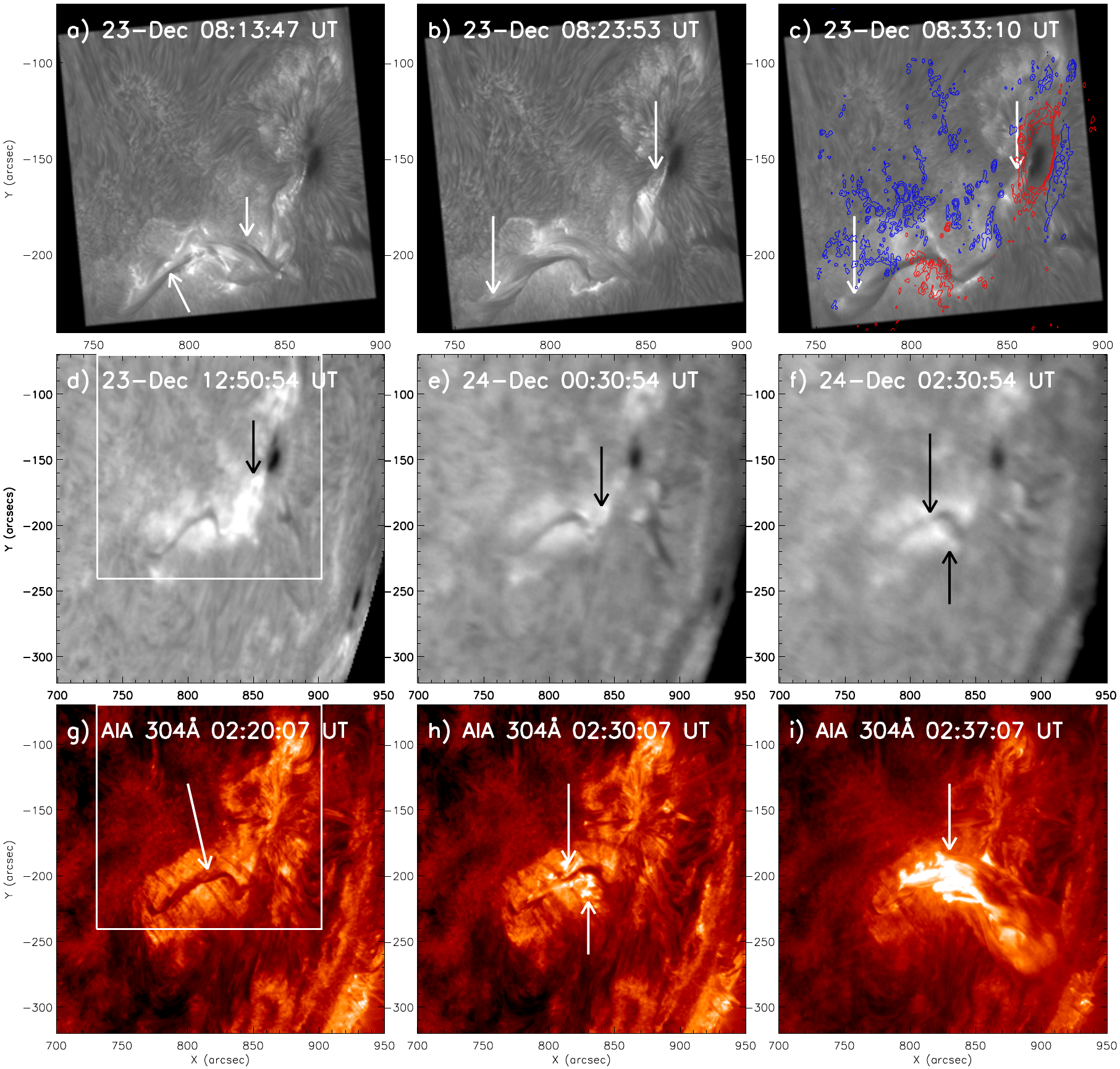}
\caption{(a-c) The filament activations in NVST H$\alpha$ filtergrams. The activated strands and the filament footpoints are indicated by the white arrows in (a) and (b-c), respectively. The contours of HMI magnetogram are overlaid in (c) with positive (negative) fields in red (blue), and the levels are 200, 400, and 600 G, respectively. (d-f) The filament activations in GONG H$\alpha$ filtergrams. (g-i) The filament evolution in AIA 304~{\AA}. The arrows in panels (d-f) and (h) show the brightenings, and the arrow in panel (g) points out the stable filament before the eruption. The arrow in panel (i) indicates the associated flare, and the boxes in panels (d) and (g) show the FOV of panels (a-c).
\label{f2}}
\end{figure}

\clearpage

\begin{figure}
\epsscale{0.9} \plotone{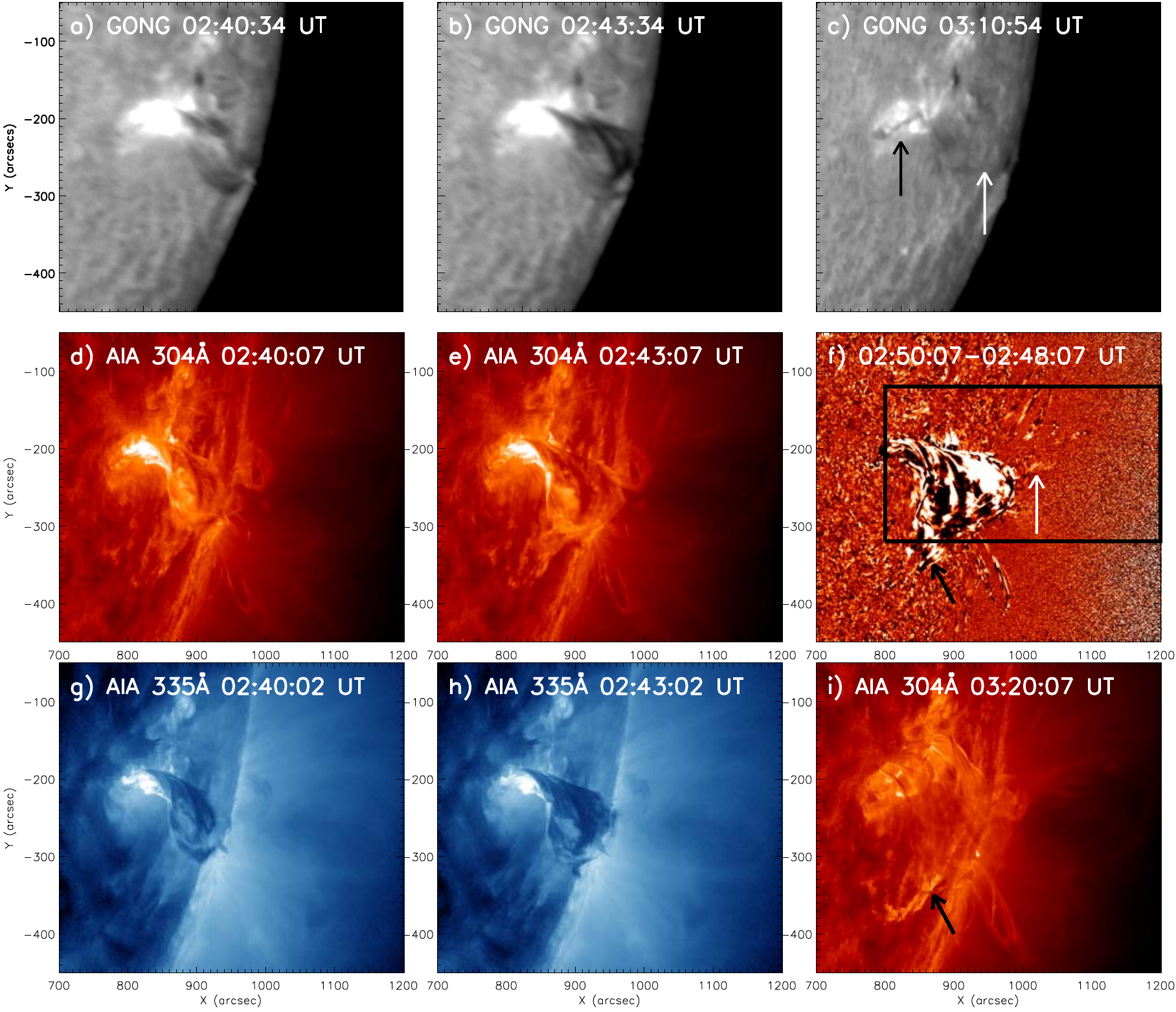}
\caption{The filament eruption in GONG H$\alpha$ filtergrams (a-c), and in AIA 304 (d-f) and 335~{\AA} (g-i). The black and white arrows in panel (c) show the confined filament material and a tiny filament, respectively. The black and white arrows in panel (f) indicate the moving plasma and the escaping blob, and the black arrow in panel (i) points to the quasi-circular flare ribbon.
\label{f3}}
\end{figure}

\clearpage

\begin{figure}
\epsscale{0.8} \plotone{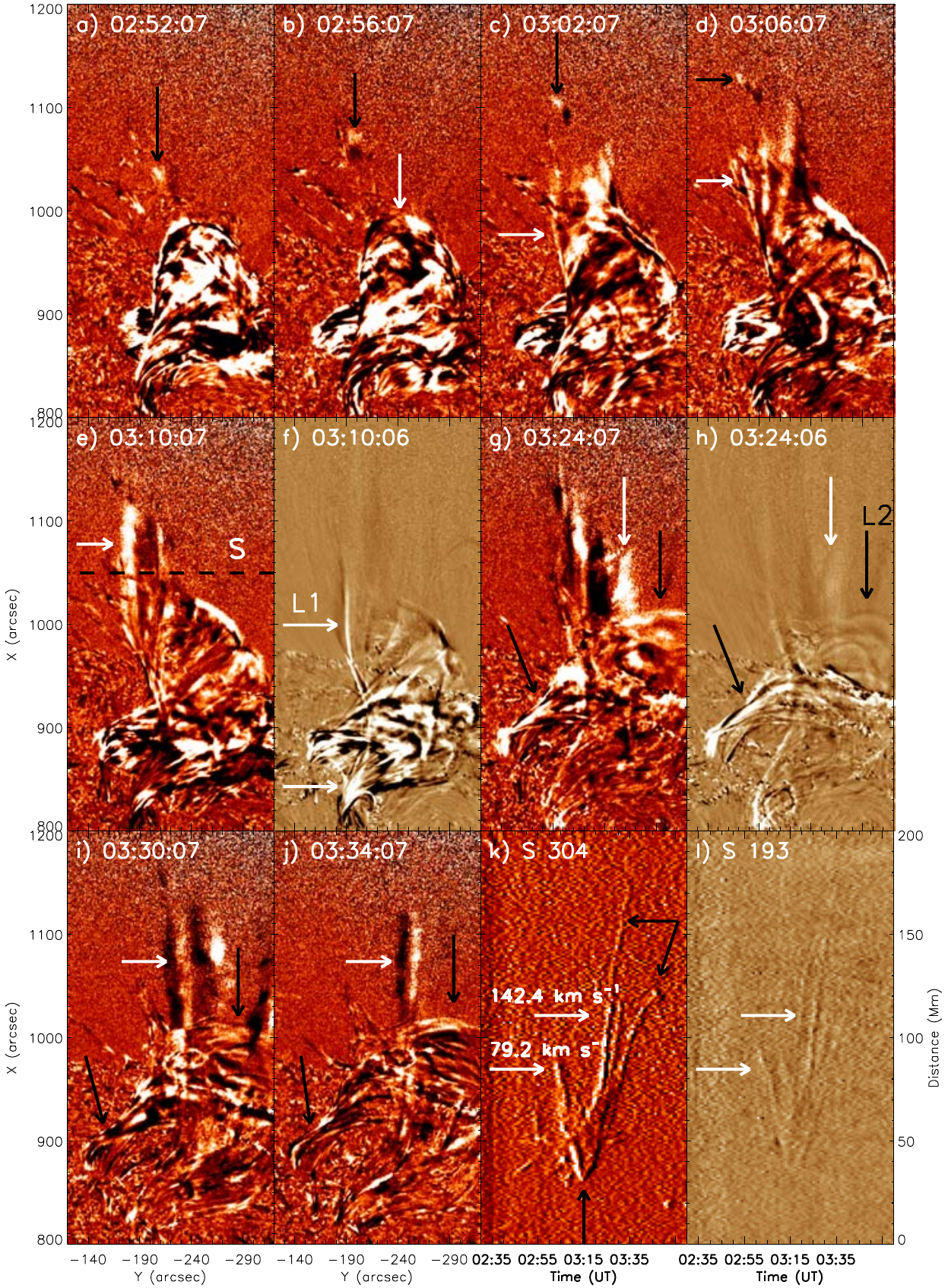}
\caption{(a-j) The evolution of the blob (black arrows in (a-d)) and newly-formed open field lines (L1; white arrows in (c-j)) in running-difference images in AIA 304 and 193~{\AA}. The white arrow in panel (b) indicates cracks of the filament body, and black arrows in panels (g-j) show the newly-formed closed loops (L2). (k-l) The time-slice plots along the dashed line (S in panel (e)). The black arrows indicate the different L1 branches, and the white arrows show that L1 first moved northward and then southward.
\label{f4}}
\end{figure}

\clearpage

\begin{figure}
\epsscale{0.9} \plotone{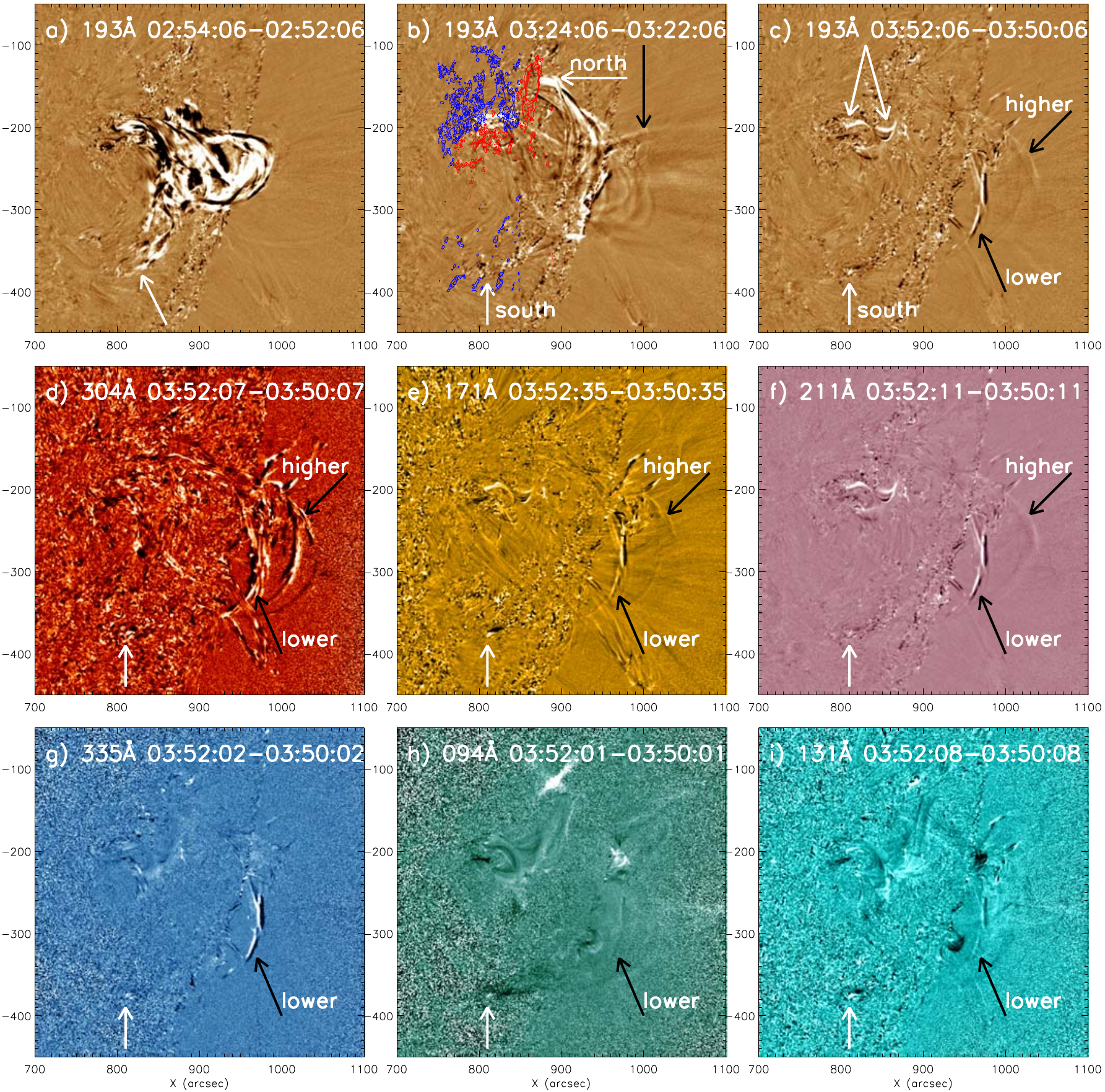}
\caption{The newly-formed closed loop (L2) in running-difference images of seven AIA channels. The black arrow in panel (b) shows L1, and the rest black arrows show L2 branches at different height. All the single white arrows indicate the footpoints of L2, and the twin white arrows in panel (c) indicate the post-eruption loops straddling the PIL. The contours of HMI magnetogram are overlaid in (b) with positive (negative) fields in red (blue), and the levels are 50, 200, 350, and 500 G, respectively.
\label{f5}}
\end{figure}

\clearpage

\begin{figure}
\epsscale{0.9} \plotone{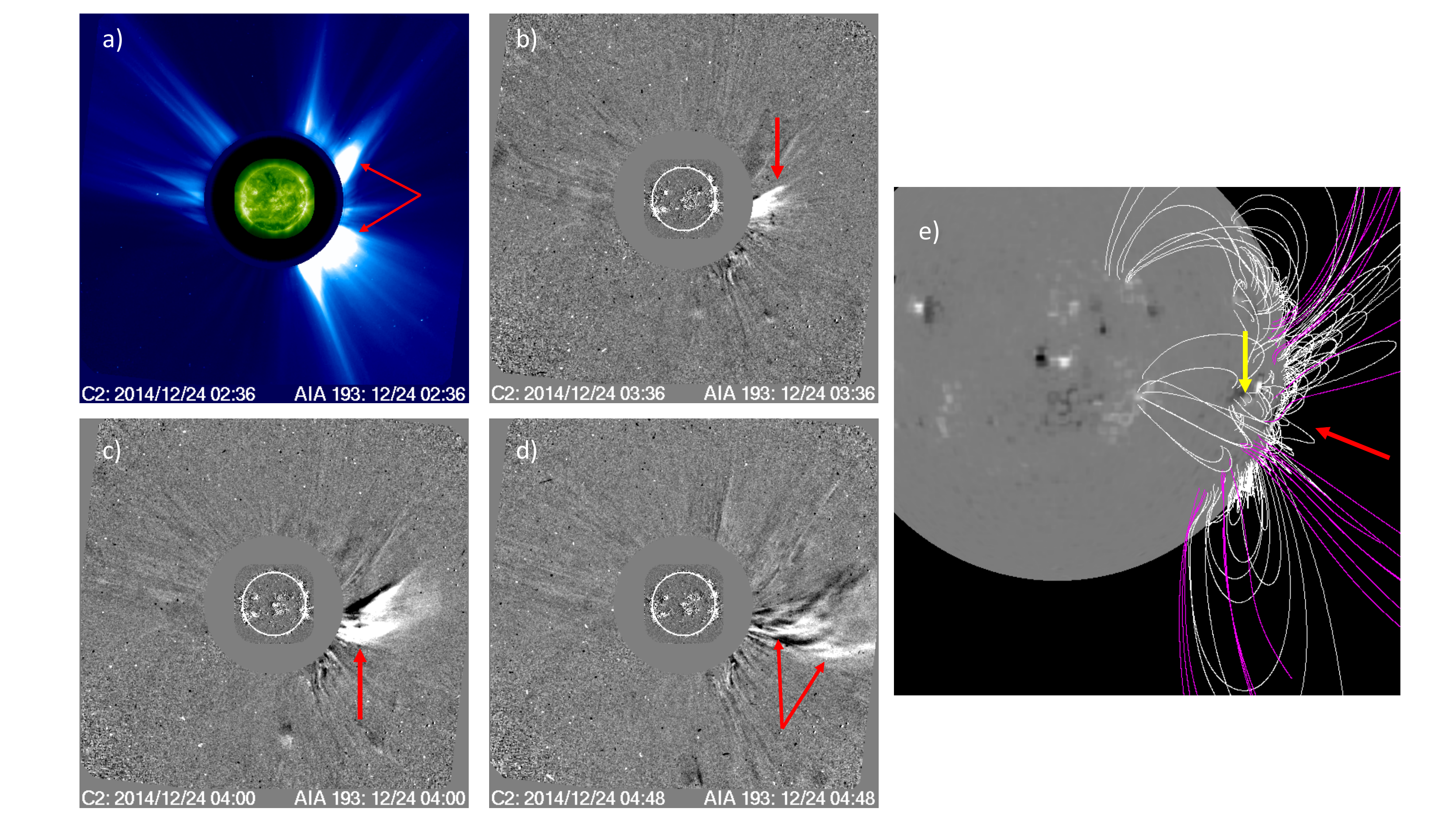}
\caption{(a-d) The nearby streamers (red arrows in (a)) and the puff-like CME (red arrows in (b-d)). (e) The PFSS results of the AR magnetic field lines, showing the nearby open field lines (pink lines). The yellow and red arrows show the location of AR and the emanating site of the CME, respectively.
\label{f6}}
\end{figure}

\clearpage

\begin{figure}
\epsscale{0.9} \plotone{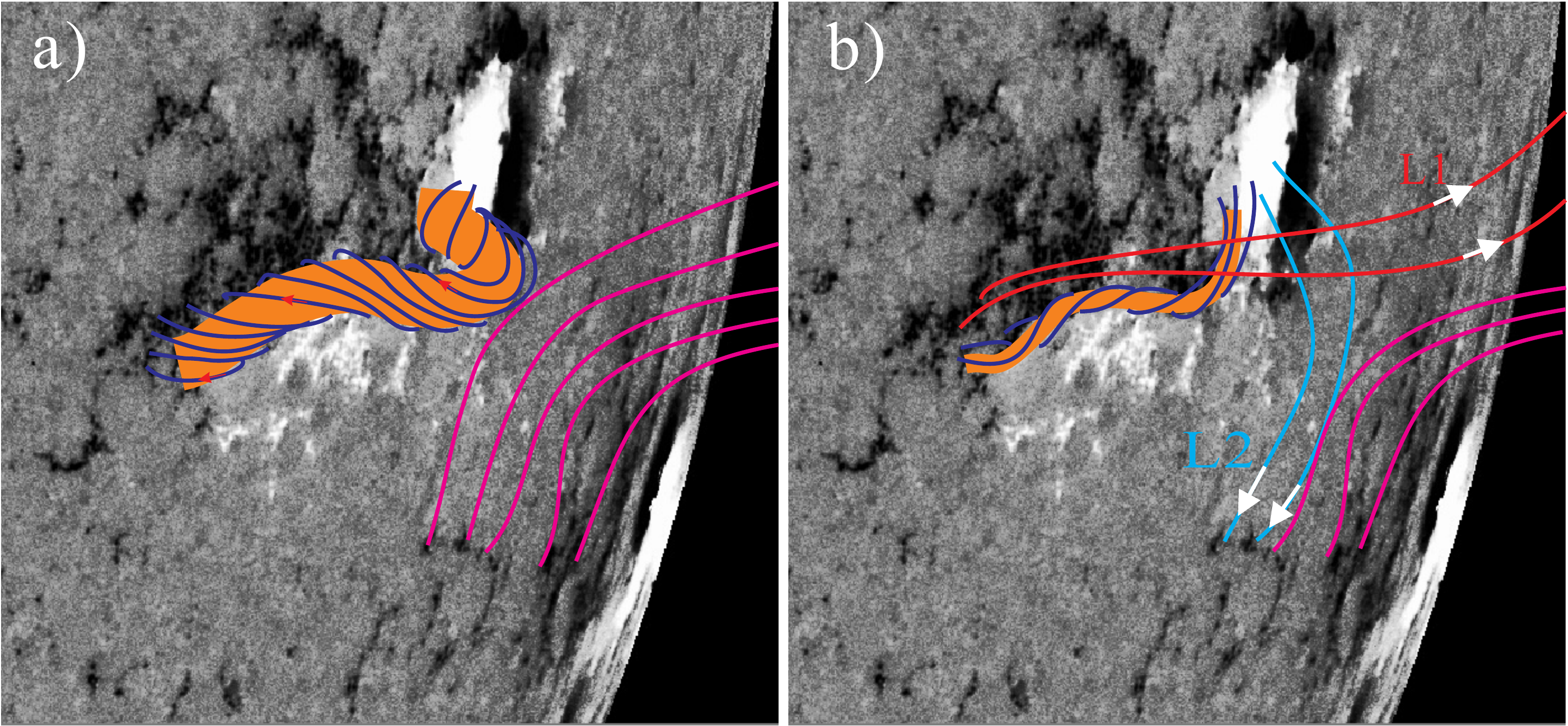}
\caption{The schematic showing the scenario of the IR between the closed field lines (deep blue) composing the filament (yellow) and the nearby open field lines (pink). After the IR, there formed new open field lines (red) and new closed loops (light blue), with blobs and bright plasma moving outward along them, respectively (white arrows).
\label{f7}}
\end{figure}

\end{document}